\documentclass[]{cjaa}

\usepackage{graphicx}                   
\input{epsf.sty}                        
\input{psfig.sty}                       

\begin{document}

   \title{Nature of giant pulses in radio pulsars}

   \volnopage{Vol.0 (200x) No.0, 000--000}      
   \setcounter{page}{1}          

   \author{S. A. Petrova}
      \inst{}\mailto{}

   \offprints{S. A. Petrova}                   

   \institute{Institute of Radio Astronomy, 4, Chervonopraporna Str., Kharkov 61002, Ukraine \\
             \email{petrova@ira.kharkov.ua}}

   \date{Received~~2005 month day; accepted~~2005~~month day}

   \abstract{Formation of giant radio pulses is attributed to
   propagation effects in the plasma of pulsar magnetosphere.
   Induced scattering of radio waves by the plasma particles is
   found to lead to an efficient redistribution of the radio
   emission in frequency. With the steep spectrum of pulsar
   radiation, intensity transfer between the widely spaced
   frequencies may imply significant narrow-band amplification of
   the radiation. This may give rise to giant pulses.
   It is demonstrated that the statistics of giant
   pulse intensities observed can be reproduced if one take into
   account pulse-to-pulse fluctuations of the plasma number
   density and the original intensity. Polarization properties of the
   strongly amplified pulses, their location in the average pulse
   window and the origin of the nanostructure of giant pulses are
   discussed as well.
   }

   \authorrunning{S. A. Petrova }
   \titlerunning{Nature of giant pulses in radio pulsars}

   \maketitle

\section{Main features of giant pulses}
\label{sect:intro}
The intensities of pulsar radio emission are known to vary from
pulse to pulse. Typically they fluctuate within a few times the
average and are believed to have Gaussian distribution. However,
several pulsars exhibit a pronounced excess of strong pulses with
a characteristic power-law distribution. These so-called giant
pulses have intensities from a few dozens up to thousands times
the average. At present there is no evidence for a break of the
power-law tail at high intensities, so the observations of longer
duration would reveal the pulses of even higher intensities. For
low enough intensities, the role of noise is significant, and the
observations do not give direct evidence as to whether giant
pulses present a separate group or whether there is a smooth
transition from normal to giant pulses. The power-law index of
giant pulse distribution differs markedly for different pulsars
and depends on radio frequency.

For a given pulsar, giant pulses constitute only a tiny fraction
of the total of pulses, less than one per cent, and do not affect
the average radio emission characteristics. The giant pulse
profiles are characterized by a specific form, with fast rise and
exponential decay, and are usually narrower than a normal pulse,
the strongest ones tending to be the narrowest.

Giant pulses occur over a broad range of radio frequencies, from a
few dozens MHz to a few GHz, but the normalized correlation
bandwidth is not large, $\Delta\nu/\nu\sim n\cdot 0.1$. Typically
giant pulses are not simultaneous out of this range (e.g. Popov \&
Stappers 2003). In case of the Crab pulsar, giant pulses are seen
simultaneously over a broad frequency range (e.g. Sallmen et al.
1999), but their substructure is correlated only over the interval
$\Delta\nu/\nu\la 0.2$ (Eilek et al. 2002).

The presence of a substructure is one of the most prominent
features of giant pulses. In the Crab pulsar, the giant pulses
often consist of several components - the nanopulses - with
timescales down to a few nanoseconds (Hankins et al. 2003). For
another well-studied source of giant pulses, PSR B1937+21, there
is also indirect evidence in favour of the nanostructure (Popov \&
Stappers 2003). Such tiny timescales imply tremendous brightness
temperatures, $10^{37}-10^{39}$ K (Hankins et al. 2003; Soglasnov
et al. 2004), making nanopulses the brightest pulses in the
Universe. Note that their energetics is only marginally consistent
with the total energetics of pulsar magnetosphere and challenges
our understanding of pulsar physics.

Similarly to the giant pulses on the whole, the nanopulses are
characterized by fast rise and exponential decay. The plot of the
peak intensity versus width has a clear upper envelope, which
corresponds to the constant integrated intensity of the nanopulses
(Eilek et al. 2002).

On a qualitative level, apart from the scaling factors, the
nanostructure of giant pulses looks very similar to the
microstructure observed in the normal pulses, hinting at a similar
origin. For a given pulsar, the microstructure shows a range of
timescales with a definite upper cutoff. The most important
property of the microstructure is that its characteristic
timescale is exactly proportional to the pulsar period (Popov et
al. 2002). The same trend can also be expected for the
nanostructure of giant pulses: The period of PSR B1937+21 is about
20 times less than that of the Crab, so its nanopulses may be
considerably shorter and, perhaps, just for this reason they still
escape direct detection.

Another remarkable phenomenon relative to giant pulses is giant
micro pulses found in the Vela and some other pulsars. Their peak
intensities are up to 40 times the peak of the average profile and
130 times the average intensity at their longitude, and they also
exhibit power-law statistics (Johnston et al. 2001). However,
giant micropulses are so narrow that their integrated intensities
are no more than a few times that of the average profile. It is
interesting to note that at certain pulse longitudes, not active
as to giant micro pulses, the intensity distribution also exhibits
a tail, though weaker than the power-law tail of giant micro
pulses. It appears consistent with log-normal statistics (Cairns
et al. 2001). Thus, even at so called non-active longitudes, the
radio emission of the Vela can be subject to amplification.

The results of polarization studies of giant pulses and relative
phenomena are not very numerous. However, on the whole one can
conclude that giant pulses, giant micro pulses and ordinary micro
pulses are strongly polarized, much stronger than the typical
normal pulses. In addition to high linear polarization,
substantial amounts of circular polarization are sometimes
present.

It has been noticed that among the known pulsars the first four
sources of giant pulses have the largest magnetic field at the
light cylinder, about a million Gauss. Moreover, this seems a
single common feature of the sources, which have very different
parameters and evolution histories. But the subsequent attempts to
discover another giant pulse sources among the pulsars with the
largest light cylinder magnetic field were not very fruitful.
Instead, a couple of giant pulse sources were discovered at low
enough frequencies, 40 \& 111 MHz (PSR B1112+50 (Ershov \& Kuzmin
2003) \& B0031-07 (Kuzmin et al. 2004)). The giant pulses are very
similar to the previously known, but the pulsars themselves have
absolutely ordinary characteristics. An impression may arise that
at low frequencies giant pulse phenomenon is so pronounced that no
additional conditions, such as high light cylinder magnetic field,
are necessary. At the same time, the actual frequency dependence
of giant pulse efficiency remains ambiguous. PSR B1133+16 presents
a counter example, showing an excess of strong pulses at high
frequencies (Kramer et al. 2003).

Despite these recent detections, giant pulses are usually
associated with the light cylinder. In particular, they are
thought to be related to the pulsar high-energy emission, whose
luminosities are also related to the light cylinder magnetic
field. An idea of such a relation has even deeper roots. Normally,
the energetics of radio emission is only a tiny fraction of the
rotational energy loss of the pulsar and the role of radio
emission processes in the global pulsar electrodynamics is
negligible. However, as giant pulses show amplifications up to a
few thousand and if they are sufficiently broad-band, their
energetics may be at a level of global electrodynamic processes in
pulsar magnetosphere. If any, the global electrodynamic changes
during giant pulses should necessarily be reflected in the
high-energy emission. However, the gamma-ray emission accompanying
giant pulses of the Crab pulsar does not show any marked changes
(Lundgren et al. 1995), the optical pulses are only 3\% brighter
(Shearer et al. 2003). In a word, the present high-energy data
testify against global electrodynamic changes during giant pulses.

Moreover, the changes in the radio emission mechanism during giant
pulses are also strongly restricted. In the Crab pulsar, the
arrival times of giant pulses show no delay with respect to the
normal ones, so that the emission region is the same (Lundgren et
al. 1995). In PSR B1937+21, during giant pulses the radio emission
beyond the narrow giant profile is also present and shows exactly
the same characteristics as in the normal pulses. One can conclude
that neither global electrodynamic processes nor special radio
emission mechanism are necessary to explain giant pulse
phenomenon. In this situation, it is reasonable to apply to
propagation effects as the radio beam passes through the plasma of
pulsar magnetosphere.

The peculiarities of giant pulse sources as to the average radio
emission characteristics have not attracted much attention yet.
But note that the largest known average radio luminosities and the
steepest known radio spectra are both the attributes of giant
pulse sources.

This is a rough observational picture of giant pulses and relative
phenomena. It contains a lot of puzzles for theorists. The most
important problems concern the source of energy supplying giant
pulses, the mechanism of energy release providing the power-law
statistics of giant pulse intensities, and the origin of the
nanostructure. All these problems can be solved in terms of
propagation effects in pulsar magnetosphere (Petrova 2004a,b).

\section{Induced Compton scattering in pulsar plasma}
Pulsar magnetospheres are believed to contain the strongly
magnetized electron-positron plasma, which streams along the open
magnetic lines with ultrarelativistic velocities. The
Lorentz-factors of the plasma $\gamma\sim 100$. The radio emission
originates deep inside the open field line tube, and further on it
should propagate through the plasma flow. The brightness
temperatures of the normal pulsar radio emission are also
extremely high, $10^{25}-10^{30}$ K, and one can expect that the
induced scattering of radio photons off the plasma particles is
strong enough to have marked observational consequences.

Induced scattering is most efficient deep inside the
magnetosphere, where the number density of the scattering plasma
particles is the largest. In this region, the approximation of a
superstrong magnetic field is valid: The radio frequency in the
particle rest frame, $\omega^\prime\equiv\omega\gamma
(1-\beta\cos\theta)$ (where $\beta$ is the particle velocity in
units of $c$ and $\theta$ the photon tilt to the ambient magnetic
field), is much less than the electron gyrofrequency,
$\omega_H\equiv\frac{eB}{mc}$. Thus, we come to the problem of
induced scattering in the superstrong magnetic field.

In the rest frame of the scattering particles, the photon
frequency is almost unaltered in the scattering act. In the
laboratory frame, this implies a certain relation between the
photon frequencies and orientations in the initial and final
states: $\omega_a\gamma
(1-\beta\cos\theta_a)=\omega_b\gamma(1-\beta\cos\theta_b)$. At a
fixed frequency, the photons are concentrated into a narrow beam
with the opening angle $\sim 1/\gamma$, but at substantially
different frequencies the beams can have different orientations
with respect to the local magnetic field. Although in the open
field line tube the wavevector tilt to the magnetic field is not
large, typically a few degrees, for different frequencies it can
differ by a factor of a few. In particular, the necessary
condition for induced scattering between the two beams of widely
spaced frequencies can be satisfied.

In the problem considered, the induced scattering transfers the
photons mainly from the low-frequency state to the high-frequency
one. With the decreasing spectrum of pulsar radiation, this may
imply a substantial amplification at the higher frequency. The
low-frequency photons are much more numerous, and if a substantial
fraction of them is transferred to the higher frequency, the
higher-frequency intensity can increase drastically. Note that
steeper radio spectra favour stronger amplification.

As long as the number of the low-frequency photons is not
negligible, the high-frequency intensity grows exponentially,
$I_a=I_a^{(0)}\exp (\Gamma)$. According to the numerical
estimates, for the giant pulse sources, the optical depth to
induced scattering is not large on average,
$\langle\Gamma\rangle\sim 0.1$, i.e. on average this process is
inefficient. Note that most of the observed pulses are indeed not
giant. However, the scattering depth depends on the number density
of the scattering particles, $N$, and the original intensity of
radiation, $I^{(0)}$, which are believed to vary randomly from
pulse to pulse. As from time to time they acquire the values more
than three times the average, the optical depth becomes large and
the intensity is strongly amplified. An exact mathematical
treatment shows that if the plasma density and original intensity
are Gaussian random variables, the final distribution of the
strongly amplified intensities is close to the power law (Petrova
2004a).

\section{Observational consequences of induced scattering in pulsars}
\subsection{Statistics of Intensity Amplification}
The numerical simulations in the framework of the amplification
model suggested, $I=I^{(0)}\exp (\Gamma)=I^{(0)}\exp \left [
\langle\Gamma\rangle\left (\frac{I^{(0)}}{\langle
I^{(0)}\rangle}\right )\left (\frac{N}{\langle N\rangle}\right
)\right ]$, allow to reproduce the observed intensity
distributions of giant pulses. Figure 1 shows the calculated
histogram of strongly amplified intensities. It is close to the
power law with the index $\alpha=-3.26$ and is similar to that
observed for the Crab pulsar at the frequency of 800 MHz (Lundgren
et al. 1995). The simulated histograms of the logarithm of
intensity for a set of average scattering depths,
$\langle\Gamma\rangle$, are presented in Fig.~2. The curves appear
close to the original Gaussian distribution, to the log-normal one
and to the distribution with the power-law tail. All types of the
histograms are indeed met in observations. In the Vela pulsar, at
different pulse longitudes both the power-law and log-normal forms
are met, testifying to the difference of the average scattering
depth at different locations in the magnetosphere. Perhaps, just
for this reason giant micro pulses are observed in this pulsar,
instead of common giant pulses.

\begin{figure}
   \centering
   \resizebox{0.7\textwidth}{!}                
            {\includegraphics[]{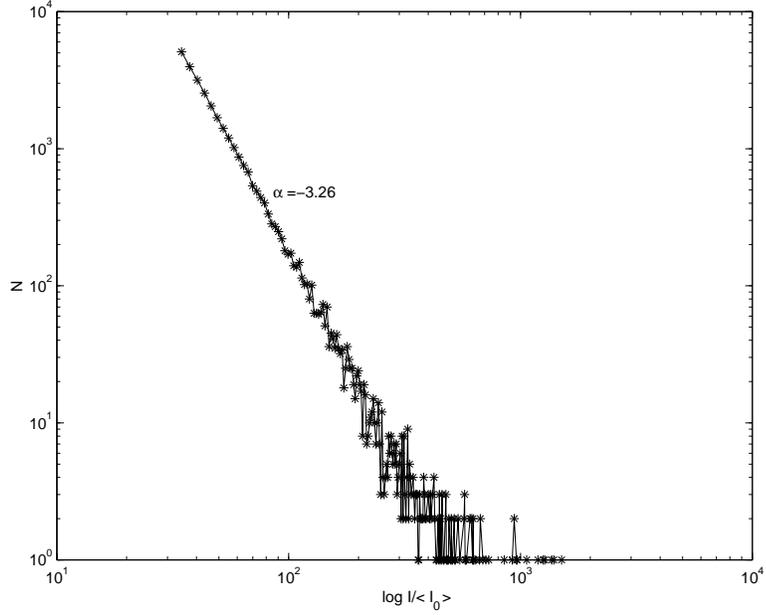}}    
   \caption{Numerically simulated histogram of strongly amplified
   intensities, $I/\langle I_0\rangle >33\langle I_0\rangle $.}
   \label{Fig:1}
   \end{figure}

   \begin{figure}
   \centering
   \resizebox{0.7\textwidth}{!}                
            {\includegraphics[]{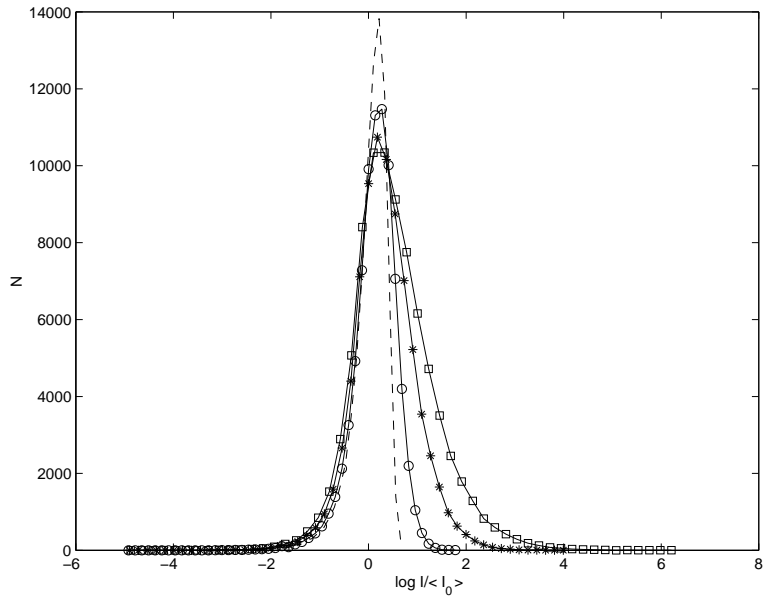}}    
   \caption{Distribution of the logarithm of intensity for different
   average scattering depths. The circles, asterisks and squares correspond to
   $\langle\Gamma\rangle =0.1$, $0.3$ and $0.5$, respectively. The dotted
   line shows the original Gaussian intensity distribution.}
   \label{Fig:2}
   \end{figure}

\subsection{Basic Characteristics of Giant Pulses}
Keeping in mind the exponential form of the intensity growth,
$I=\langle I_0\rangle\exp (\Gamma)$, one can estimate the
bandwidth of the amplified radiation as $\Delta\nu/\nu\sim
1/\Gamma$. Given that the maximum observed intensity amplification
is about a few thousand times, the maximum scattering depth should
be approximately 7. Hence, the estimated bandwidth of giant
radiation roughly corresponds to the observed value of a few
tenths. It should be noted, however, that the conditions for
efficient beam-to-beam scattering can be satisfied simultaneously
in a number of distinct regions in the open field line tube and
different frequencies can be amplified independently. This may
result in a broad-band activity of giant pulses similar to that
observed in the Crab pulsar.

The width of a giant pulse can be roughly estimated as $w_{\rm
GP}\sim w/\Gamma$. Hence, giant pulses should indeed be somewhat
narrower, with the strongest ones tending to be the narrowest,
just as is really observed. Note, however, that the actual
profiles of giant pulses should be determined by the location and
geometry of the scattering region. In addition, giant pulses are
strongly affected by interstellar scattering.

The suggested mechanism of giant pulses implies a strong
polarization of the substantially amplified radiation. The
ultrarelativistic strongly magnetized plasma of pulsars allows two
non-damping natural waves - the ordinary and extraordinary ones, -
which are linearly polarized in orthogonal directions. In general,
the pulsar beam presents an incoherent superposition of the two
types of waves and is partially depolarized. In the superstrong
magnetic field, the process of induced scattering is efficient
only for the ordinary waves. Hence, only the ordinary waves are
amplified, whereas the fraction of the extraordinary ones in the
total radio emission becomes so small that they cannot cause
substantial depolarization. Thus, giant pulses should be strongly
polarized and have the polarization of the ordinary waves, with
the electric vector lying in the plane of the ambient magnetic
field. The latter is still to be confirmed by observations.

\subsection{Location of Giant Pulses in the Average Pulse Window}
The necessary condition for induced scattering between the rays of
widely spaced frequencies and orientations, i.e. the condition of
giant amplification, can be satisfied only within some specific
regions in the magnetosphere. In case of PSR B1937+21, which has
the shortest known period, $P=1.56$ ms, the rotational effect is
very strong - even the stellar surface rotates at a speed of
$0.1c$. The rays of different frequencies are thought to originate
at different altitudes above the neutron star. (Higher frequencies
are emitted deeper in the magnetosphere, where the plasma number
density and the characteristic plasma frequency are larger.) The
rays emitted at different altitudes propagate in the rotating
magnetosphere and travel to the point of scattering for different
time intervals. So they indeed come at substantially different
angles.

An exact geometrical consideration leads to the conclusion that
the necessary condition for induced scattering between the widely
spaced frequencies can be satisfied only for the rays emitted
almost parallel to the magnetic axis in the laboratory frame
(Petrova 2004a). Because of rotational aberration, the rays
co-directed with the magnetic axis appear in the trailing part of
the pulse profile. The giant pulses of PSR B1937+21 are indeed met
within a narrow window at the trailing edge of the profile (e.g.
Soglasnov et al. 2004). The observed time delay from the center of
the average profile, $\tau\sim 60\,\mu$s, translates to the
aberration angle $\Delta\theta=2\pi\tau /P\approx 0.2$, which
seems quite realistic. Thus, it is very likely that the location
of giant pulses in the profile of PSR B1937+21 coincides with the
projection of the magnetic axis and is determined by the
rotational effect.

As for the Crab pulsar, its period is about 20 times longer, and
perhaps the rotational effect is too weak to provide the necessary
geometrical condition of induced scattering. In this case, the
necessary condition can be set up by the frequency-dependent
refraction of radio waves. The location of the scattering region
is then determined by the instantaneous distribution of the plasma
in the open field line tube, which can markedly fluctuate, so that
giant pulses can be met at different locations within the average
profile.

\section{Origin of the micro/nano structure}
The nanostructure of giant pulses and the relative phenomenon of
microstructure in the normal pulsars can also be explained in
terms of propagation effects in pulsar plasma.

The proportionality of the characteristic timescale of
microstructure to pulsar period strongly supports the angular
beaming model: The micro/nano pulses result from the inhomogeneity
of the angular structure of pulsar beam, which manifests itself as
the pulsar rotates. Then the angular scale of inhomogeneity is the
same for all pulsars, while the difference of the timescales
results simply from the difference in the velocity of rotation
with respect to an observer, $\Delta\tau=P\Delta\theta/(2\pi)$.
The angular scale of inhomogeneity is commonly associated with the
opening angle of relativistically beamed radiation,
$\Delta\theta\sim 1/\gamma$. However, this requires huge
Lorentz-factors of the plasma, $10^4$ for the microstucture and
$10^7$ for the nanostructure, while the modern theories of pair
creation cascade in pulsars give the values $\gamma\sim 10^2$.
Besides that, it is difficult to explain a wide range of the
timescales met in a given pulsar.

These difficulties can be avoided if one take into account induced
scattering inside the photon beam (Petrova 2004b). Let a narrow
beam, with an opening angle $\Delta\theta\sim 1/\gamma$, propagate
through the plasma flow at not a small angle to the magnetic
field, $\theta\gg 1/\gamma $. The induced scattering acts mainly
to redistribute the photon orientations inside the beam. The
photons tend to be focused in the direction closest to the
magnetic field direction.

Because of the focusing effect, the beam can be substantially
squeezed, its angular width is now
$\Delta\theta\sim1/(\gamma\Gamma_b)$, where $\Gamma_b$ is the
optical depth to the scattering inside the beam. The observed
timescales of microstructure require $\Gamma_b\sim 200$, which is
consistent with the theoretical estimates of this quantity
(Petrova 2004b). The focusing effect as a result of induced
scattering not only removes the problem of exceedingly large
Lorentz-factors of the plasma, but also explains a large scatter
of the timescales observed in a given pulsar. This scatter is
naturally attributed to the fluctuations in the plasma flow, in
which case the scattering efficiency and the extent of focusing
fluctuate as well. It is especially interesting to note that for a
few pulsars studied the maximum observed scale of microstructure
is compatible with the angular scale $\sim 1/\gamma$, which is
indeed the case under condition of weak scattering, when
$\Gamma_b\ll 1$.

The induced scattering inside the beam certainly holds during
giant pulses, in addition to the beam-to-beam scattering which
causes intensity amplification, and it is thought to be much more
efficient because of the larger plasma densities and initial
intensities peculiar to giant pulses. For the Crab pulsar we have
the following estimates. The average optical depth to beam-to-beam
scattering is $\langle\Gamma\rangle\sim 0.1$, whereas the maximum
value $\Gamma_{\rm max}\approx 7$. As for the scattering inside
the beam, $\langle\Gamma_b\rangle\sim 500$, and if this quantity
can also increase by 70 times, the beam width is $\Delta\theta\sim
1/(70\langle\Gamma_b\rangle\gamma)$, which corresponds to the
timescale $\Delta\tau=\Delta\theta P/(2\pi)\sim 10^{-9}$~s. Thus,
the focusing effect as a result of induced scattering inside the
beam can explain the structural details of giant pulses at
timescales down to a nanosecond. Note that this interpretation
naturally removes the problem of exceedingly high brightness
temperatures of the nanopulses.

An idea of the focusing effect as the origin of the nanopulses
finds a certain support in the observational plot of the peak
intensities of the nanopulses versus their width, which has a
clear upper envelope corresponding to the constant integrated
intensity (Eilek et al. 2002). The points along the envelope can
result from the focusing of different extent because of induced
scattering inside the beam at slightly different physical
conditions, whereas the points below the envelope can correspond
to different levels of true amplification due to the beam-to-beam
scattering.

\section{Conclusions}
The main features of giant pulses and relative phenomena can be
explained in terms of propagation effects in pulsar magnetosphere.
Giant pulses can be caused by induced scattering in the flow of
pulsar plasma. Induced scattering between substantially different
frequencies and orientations leads to a marked redistribution of
radio intensity in frequency. In the approximation of a
superstrong magnetic field, the radio photons are mainly
transferred from the lower-frequency state to the higher frequency
one. With the decreasing spectrum of pulsar radiation, this may
lead to a substantial increase of the higher-frequency intensity.

The optical depth to the beam-to-beam scattering is not large on
average, but it depends on the plasma number density and the
original radio intensity, which are believed to vary randomly from
pulse to pulse. As from time to time they become larger than three
times the average, the scattering depth becomes significant and
amplification is efficient. Given that the plasma number density
and the original intensity are Gaussian random variables, the
distribution of substantially amplified intensities is close to
the power-law. Numerical simulations in the framework of the
suggested model of intensity amplification allow to reproduce the
observed intensity distributions of single pulses.

The mechanism suggested explains the location of giant pulses in
the average pulse window in cases of the Crab and PSR B1937+21 and
predicts strong polarization of giant emission in the plane of the
ambient magnetic field. The nanostructure of giant pulses, as well
as the microstructure of the normal ones, are explained by the
focusing effect as a result of induced scattering inside the beam.
The problem of exceedingly large brightness temperatures of the
nanopulses is removed naturally.

\begin{acknowledgements}
I am grateful to the Organizing Committee for the financial
support which has made possible my participation in the Symposium.
The work is in part supported by INTAS Grant No.~03-5727.
\end{acknowledgements}

\label{lastpage}

\end{document}